\documentclass[letterpaper]{article}
\usepackage{jheppub_mod}
\pdfoutput=1
\usepackage{graphicx, caption, subcaption}  
\usepackage{dcolumn}   
\usepackage{amsmath, amssymb}        
\usepackage{color,latexsym, array,multirow,  verbatim, enumerate}
\usepackage{url}
\usepackage[normalem]{ulem}

\usepackage{enumitem}
\usepackage{bbding}
\usepackage{tikz}
\usepackage{pifont}

\newcommand{\be}{\begin{equation}}
\newcommand{\ee}{\end{equation}}
\newcommand{\bea}{\begin{eqnarray}}
\newcommand{\eea}{\end{eqnarray}}
\newcommand{\ba}{\begin{eqnarray*}}
\newcommand{\ea}{\end{eqnarray*}}
\newcommand{\barr}{\begin{array}}       
\newcommand{\earr}{\end{array}}

\newcommand{\Eq}[1]{Eq.~(\ref{#1})}
\newcommand{\Sec}[1]{Sec.~\ref{#1}}
\newcommand{\Fig}[1]{Fig.\,\ref{#1}}
\newcommand{\Tab}[1]{Table~\ref{#1}}
\newcommand{\Cit}[1]{~\cite{#1}}

\newcommand{\tev}{~\text{TeV}}
\newcommand{\gev}{~\text{GeV}}

\newcommand{\circleb}[1]{\tikz \node[circle,inner sep=0.5 pt,color=white, fill=black]{\footnotesize #1};}
\newcommand{\circlew}[1]{\tikz \node[circle,draw, inner sep=0.5 pt,color=black, fill=white]{\footnotesize #1};}

\title{Augmenting Collider Searches and Enhancing Discovery Potentials through Stochastic Jet Grooming}

\author[a]{Tuhin S.~Roy, }
\author[b,c]{and Arun M.~Thalapillil}

\affiliation[a]{Department of Theoretical Physics, Tata Institute of Fundamental Research,\\
  Mumbai 400005, India}
\affiliation[b]{NHETC, Department of Physics and Astronomy,\\
Rutgers University,\\
Piscataway, NJ 08854}
\affiliation[c]{Indian Institute of Science Education and Research,\\
Homi Bhabha Rd, Pashan, Pune 411 008, India}

\emailAdd{tuhin@theory.tifr.res.in}
\emailAdd{thalapillil@iiserpune.ac.in}

\date{\today}
\abstract{The jet Trimming procedure has been demonstrated to greatly improve event reconstruction in hadron collisions, by mitigating contamination due initial state radiation, multiple interactions, and event pileup. Meanwhile, Qjets -- a nondeterministic approach to tree-based jet substructure has been shown to be a powerful technique in decreasing random statistical fluctuations, yielding significant effective luminosity improvements. This manifests through an improvement in the significance $S/\delta B$, relative to conventional methods. Qjets also provide novel observables in many cases, like mass-volatility, that could be used to further discriminate between  signal and background events. The statistical robustness and volatility observables, for tagging, are obtained simultaneously. We explore here a combination of the two techniques, and demonstrate that significant enhancements in discovery potentials may be obtained in non-trivial ways. We will illustrate this by considering a diboson resonance analysis as a case study -- enabling us to interpolate between scenarios where the gains are purely due to statistical robustness and scenarios where the gains are also reinforced by volatility variable discriminants. The former, for instance, is applicable to digluon/diquark resonances, while the latter will be of relevance to di-$W^\pm$/di-$Z^0$ resonances, where the boosted vector bosons are decaying hadronically and have an intrinsic mass scale attached to them. We argue that one can enhance signal significance and discovery potentials markedly through stochastic grooming, and help augment studies at the Large Hadron Collider and future hadron colliders.}

\keywords{LHC, Jets, Jet substructure, Jet grooming, Trimming, Qjets, Volatility, Diboson resonance, Dijet resonance, QCD jets.}

\preprint{}

\begin{document}

\maketitle
\flushbottom

\section{\label{sec:intro}Introduction}
The Large Hadron Collider (LHC) has spectacularly confirmed our understanding of the Standard Model (SM) of particle physics, primarily among them through the discovery of the long awaited Higgs boson\Cit{{Chatrchyan:2012ufa},{Aad:2012tfa}}. The LHC is currently in its Run-II phase and various signal topologies are being actively searched for, further expanding our quest for physics Beyond Standard Model (BSM).  Unfortunately,  the absence, so far, of any persistent and unambiguous BSM signals at the LHC (as well as related low-energy precision/flavor experiments) has put some of the most popular scenarios like Composite Higgs, Extra Dimensions and low-energy Supersymmetry under tension (See for \textit{e.g.}\Cit{Craig:2013cxa}, for a nice summary of the latter's status after Run I).  In the complete absence of any hint for  `well-motivated' BSM scenarios, the importance of general purpose searches cannot be emphasised enough. 

Possibly, the  simplest example of such general purpose searches is the search for heavy resonances decaying into pairs of reconstructed objects -- such as  isolated-photons, leptons or jets.   Indeed, the recent excitement in particle physics community arose from one such search, namely the search for a di-photon resonance (see\Cit{Strumia:2016wys} and references therein). There may even be tantalising hints of new physics in other diboson searches\Cit{{Khachatryan:2014hpa},{Aad:2015owa}}, yet to be confirmed and understood. On one hand, this general strategy is sensitive to varied models and topologies, and on the other hand, is unavoidably plagued sometimes by the lack of sufficient handles in the event. Since event information cannot be used, often times the scope of these methods are limited by the efficacy of identifying/tagging reconstructed objects, and the invariant mass resolution of the resonances. Clearly, these generic searches are more likely to succeed when the decay products contain hard photons and leptons, and less so when decay remnants only give rise to jets, because of the ubiquitous SM backgrounds (mostly due to QCD).  

If the hadronically decaying resonance is significantly boosted, such that its decay products remain tightly collimated, it would in turn give rise to a single \emph{fat-jet}; in these cases tackling background becomes relatively easier. 
Techniques based on jet substructure physics have enabled us to effectively tag jets arising from the decay of boosted heavy particles,  improve searches for new signal topologies, investigate various jet properties and mitigate contamination\Cit{Seymour:1993mx, Butterworth:2007ke, Brooijmans:1077731,Butterworth:2008iy, Butterworth:2009qa, Kaplan:2008ie,Thaler:2008ju,Almeida:2008yp, Butterworth:2008sd, Butterworth:2008tr, Plehn:2009rk, Ellis:2009su, Ellis:2009me,Krohn:2009th, Thaler:2010tr, Gallicchio:2010sw, Soper:2011cr, Fan:2011jc, Englert:2011iz, Thaler:2011gf, Plehn:2011tg, Jankowiak:2011qa, Gallicchio:2011xc, Gallicchio:2011xq, Ellis:2012sd, Ellis:2012zp, Soper:2012pb,Gallicchio:2012ez, Krohn:2012fg, Soyez:2012hv, Ellis:2012sn,Larkoski:2013eya, Larkoski:2014gra, Ortiz:2014iza, Cacciari:2014jta,Cacciari:2014gra, Bertolini:2014bba, Ellis:2014eya,Pedersen:2015knf, Pedersen:2015hka,Frye:2016okc}. (See\Cit{Adams:2015hiv} and references therein for a review and comparison of some of these techniques).

Even though these tools provide impressive signal to background separation (namely, $S/B$) or discovery potential (namely, $S/\delta B$) in the boosted limit, we have made only limited progress, relatively, for the non-boosted scenarios where the heavy resonances give rise to multiple jets. Marginal improvements in  $S/B$ and $S/\delta B$ has been reported~\cite{Krohn:2009th}, if the jets that are used to reconstruct the resonance are `trimmed' first. Jet Trimming\Cit{ Krohn:2009th} was developed, with light parton jets in mind, as a means to reduce contamination from soft radiation; primarily from initial state, multiple interactions, and pileup (terms defined later). The aim was to extract from the event record, as optimally as possible, information corresponding to the hard scattering of interest. It was argued that one may achieve good improvements in signal reconstruction, in the presence of generic QCD backgrounds\Cit{ Krohn:2009th}. The method has since been convincingly demonstrated, in numerous analyses, to be a powerful tool, by both the LHC collaborations (See for instance \cite{Chatrchyan:2013vbb, LENZI:2013wta, GUETA:2013rta, LOCH:2014lla, CMS:kxa, Aad:2013gja, TheATLAScollaboration:2013ria}).   

The aim of this paper is to provide techniques that further enhance both $S/B$ and $S/\delta B$ over ordinary Trimming, and at the same time remain relatively pileup robust. In order to achieve it, we propose a stochastic grooming methodology -- QTrimming, a technique that improvises Trimming by borrowing ideas from Qjets; a nondeterministic approach to jet substructure\Cit{Ellis:2012sn}. Qjets was originally motivated by the fact that the actual parton shower is not technically invertible, with the various jet algorithms attempting to as closely approximate the $p_T$-ordered or angular-ordered shower sequence, or both. This way of interpreting jets through multiple sets of possible showering histories, was shown to enhance discovery potentials significantly. The non-trivial statistical properties of the method were further studied in\Cit{Ellis:2014eya}, demonstrating both quantitatively and through simulation, how the improvements arise. It was also emphasised that the stochastic method provides new distributions in many cases, of various jet observables, giving an additional handle in discriminating signal and background\Cit{Ellis:2012sn,Ellis:2014eya}. This methodology of multiple interpretations has since been extended to the case of $h\rightarrow b \bar{b}$ events\Cit{Kahawala:2013sba, Chien:2013kca, Chien:2014hla}, finding remarkable improvement in significance. The effectiveness of the Qjet paradigm, in real world scenarios, has been corroborated now at the LHC\Cit{TheATLAScollaboration:2013tia,CMS:2013uea,Khachatryan:2014vla, CMS:2014joa}. 

 A central theme of this paper originates from the understanding that like most groomers, Trimming depends on the clustering history. The Qjets algorithm involves assigning to a jet many clustering histories, each of which results in a distinct Trimmed jet. The groomed jet from the usual Trimming approach (henceforth referred to as conventional Trimming), can be thought of as a single member of this ensemble of trimmed jets. If a tagging requirement is made on the trimmed jet (for example, whether the jet falls within a specified mass window or not), it is clear that conventional Trimming maps a jet to a binary probability distribution (i.e. tagged or not tagged). On the other hand, in case of QTrimming, one can ask for the fraction of iterations that give rise to a groomed jet that satisfies the tagging criteria. Merging Trimming with the Qjets paradigm therefore maps a jet to a real number, in the interval $\left[0,1\right]$, designating the tagging efficiency for the jet. This will be shown to generically enhance discovery potentials, by rendering additional statistical stability to the jet observables as well as providing new handles for event discrimination. A comparison of similar methodologies, as applied to other jet grooming techniques, is beyond the scope of the present work and we delegate it to a forthcoming study.  
 
As a case study, we will demonstrate the potential for our approach by looking at a toy example, where a heavy resonance ($\Phi$) decays to $W^\pm$ bosons, which subsequently decay to jets
\begin{equation}
	p~p \ \rightarrow \ \Phi \ \rightarrow \  W\left( q\bar{q}'\right)  W\left( q\bar{q}'\right) \;. 
\label{eq:1}	
\end{equation}
Note that in the limit of interest, $m_\Phi \gg m_W$, the decay products from the $W^\pm$ will be collimated; we will refer to these collimated jets from the $W^\pm$ decays as the daughter jets. So we have the interesting and rich scenario where multiple particles are to be tagged -- $W$ tagging (say, in the simplest case, if the corresponding daughter jet-masses are in a prescribed mass window); $\Phi$ tagging (say if the reconstructed dijet mass is in a particular mass-window). For an event to be finally tagged as the topology of \Eq{eq:1}, both daughter jets need to be tagged as $W^\pm$ and the dijet, constructed from the daughter jets, need to be tagged as $\Phi$. Again, for conventional Trimming, an event is either tagged or not tagged. In contrast, for QTrimming there are various ways to map the event to a real number representing the event tagging efficiency, in non-trivial ways. For example, the event tagging efficiency could be taken as the product of the tagging efficiencies of each of the daughter jets (as candidates for $W^\pm$) and the tagging efficiency for the dijet 4-vector (as a candidate for $\Phi$). We shall explore various cases systematically and quantify the gains for each choice. As a general feature, it is found that as the  Qjets algorithm is altered to accommodate Trimming, it produces an improved, statistically stabilised, counting of background events -- \textit{i.e.} $\delta B < \sqrt{B}$, a feature characteristic of the Qjets paradigm\Cit{Ellis:2012sn,Ellis:2014eya}. An added benefit of the stochastic method is that in many cases it also provides new distributions for various jet observables (namely volatilities, defined later). These are obtained automatically, without any additional computations, and give an additional handle in discriminating signal and background\Cit{Ellis:2012sn,Ellis:2014eya}. By applying a cut on these volatilities, in relevant cases, one can hope to improve $S/B$ and  $S/\delta B$ significantly. We will utilise a simple mass-volatility variable to demonstrate this and will delineate the specific improvements due to it.

We will explore various limits of tagging an event, in the context of QTrimming, to understand in detail the various ways in which discovery potentials may be enhanced. This will also provide a unified way to look at the method as a whole. The idea will be to progressively incorporate more and more information, from the stochastic perspective, thereby furnishing a nice interpolation between cases where the enhancements are purely due to statistical robustness and cases where the gains are also augmented by volatility discriminants. These cases are quite general -- the former is applicable to any digluon/diquark resonance, and the latter to any di-$W^\pm$/di-$Z^0$ resonance, where the boosted vector bosons are decaying hadronically. Investigating various cases, we shall argue that the stochastic grooming method can significantly augment collider searches and improve discovery potentials.

In \Sec{sec:2}, to fix our terminologies and notations, we briefly review the concept of jets and the various sequential jet algorithms, along with some of the current techniques being employed to reduce contamination.  In \Sec{sec:3}, the idea of a nondeterministic approach to jet formation is briefly reviewed, emphasising the various statistical quantities of interest, and motivating the gain in statistical stability endowed by this approach.  We will describe our specific procedure in \Sec{sec:4}, and illustrate in \Sec{sec:5} that by combining Trimming with  nondeterministic, tree-based jet substructure we can achieve significant gains; much beyond improvements obtainable in the conventional, deterministic approaches. \Sec{sec:6} contains a summary of our study and the main conclusions.

\section{\label{sec:2} Jets and Jet Grooming Techniques }
In this section, we will briefly discuss some concepts related to jets and jet grooming methods, relevant to the present work, fixing terminology and notations along the way.
 \subsection{Parton Showers and Sequential Jet Algorithms}
 
Many of the searches at the LHC and the high energy frontier involve coloured final states. Due to asymptotic freedom, any hard parton produced directly or from decays, ultimately manifest as color-singlet hadrons at long distances. This process is generally viewed as a multistage, parton-shower evolution with the collection of nearby final-state hadrons being identified as a jet at the end. 

Jets are constructed by clustering various detector elements (such as calorimetric cells, tracks, or even particle flow $4$-vectors) employing a suitable jet algorithm. The first such procedure was developed in the context of $e^+ e^-$- collisions\Cit{Sterman:1977wj}. Various algorithms exist currently, differing in their methodology and clustering behaviour; though these inherently present a level of ambiguity in any jet-based measurements, any infrared safe algorithm will yield valid results that may be compared to theoretical calculations (See for e.g.\Cit{Ellis:2007ib,Salam:2009jx} and references therein). 

An important class of jet algorithms are the sequential recombination algorithms, where jets are constructed iteratively, starting from final state particle momenta. They are used prolifically in most of the current analyses. Primarily, two metrics are involved at each step of a sequential recombination jet algorithm
\begin{equation}
\begin{split}
 d_{ij}& \ = \  \text{min}\left( p_{T_i}^{2p}, p_{T_j}^{2p} \right) \frac{\Delta R_{ij}^2}{R^2},\\
 d_{i} & \ = \ p_{T_i}^{2p}\; .
 \end{split}
\end{equation}
Here, $\Delta R_{ij} = \sqrt{(y_i - y_j)^2 + (\phi_i - \phi_j)^2}$ is the angular distance between a pair of $4$-vectors $i$ and $j$ with $y$, $\phi$, and $p_T$ being the rapidity, the azimuthal angle, and the transverse momentum respectively. $R$ represents the characteristic size of the jet to be constructed. At each step, the above measures are computed and if the smallest among these correspond to a two-object measure ($d_{ij}$), the objects $i$ and $j$ are merged and taken to the next step in the sequence. If the smallest measure is a one-object measure ($d_i$), it is removed from further iterations and is designated a jet. The process is continued until all objects have been merged (an inclusive jet algorithm) or until a fixed distance between objects has been achieved (an exclusive jet algorithm). Depending on the choice of the parameter $p$, signifying our approximation of the unknown parton-shower history, the sequential algorithms have various names -- $k_T$ ($p=1$)\Cit{Catani:1993hr, Ellis:1993tq}, Cambridge-Aachen or C/A ($p=0$)\Cit{Dokshitzer:1997in, Wobisch:1998wt} and Anti-$k_T$ ($p=-1$)\Cit{Cacciari:2008gp}.
   
Note that given a choice of a sequential recombination algorithm, the tree history leading to a jet, from a set of final states, is completely deterministic. In \Sec{sec:3}, we will see that this procedure may be recast in a nondeterministic way, rendering the construction largely independent of the jet measure used and therefore more likely to characterise the physical features of a jet; as opposed to some artifact from the particular, deterministic, recombination jet algorithm being used.

\subsection{Underlying Events and Jet Grooming} 

 If a jet can be constructed as faithfully as possible to true scattering, one may hope to achieve optimal signal characterization and significance. This is generally only possible if the final states detected in a detector came solely from hard scattering, without any contamination -- in short, an idealization too far removed from reality. Apart from implicit unknowns in the parton-shower history, which we shall discuss briefly in the next section, the detector will also record final states coming from vertices other than the primary, hard-scattering vertex of interest. 
 
 The incoming initial states will generally radiate before undergoing scattering, leading very often to substantial initial state radiation (ISR). Furthermore, one has to worry about non-primary scatterings between partons, constituting the hadrons that are participating in a hard scattering. These scattering events are usually termed multiple-interactions (MI). At the LHC, where bunches of a large number of protons are made to collide at an interaction point, there is further complication due to interactions between other non-primary hadrons in the colliding bunches. This contribution to jet contamination goes by the name of pileup (PU). The above effects are especially pertinent for modern hadron colliders, like the LHC, due to their large energies and luminosities. It is crucial for new physics studies to mitigate these effects, before attempting to characterise signal-like or background-like events.
 
Thus, any jet algorithm is forced to deal with an inevitable dichotomy -- one would like to form jets large enough to include all of the hard scattering decay products and account for wide angle final state radiation (FSR), while on the other hand balance how large jets can be, due to ISR/MI/PU contamination.
 
This dichotomy may be addressed in many ways. One simple way is to try and choose an optimal jet size that achieves both objectives to the best extent possible. Another approach would be to incorporate observables that are relatively immune from contamination affects. A more active method would be to try subtract off contributions from ISR/MI/PU. In this approach one could, for instance, try to subtract off from a jet, a fixed contribution\Cit{Cacciari:2007fd}, proportional to the jet area\Cit{Cacciari:2008gn} or shape\Cit{Soyez:2012hv}. Alternatively, one could aggressively groom each jet by ``unwrapping" its constituent parts. Such an approach is motivated by the observation that there is usually just a single hard scattering, in each event, with all other sources of radiation like ISR/MI/PU stemming from much softer scatterings. 

Thus, by unwrapping a jet and removing soft radiation through a modification of the sequential clustering procedure or by analysing the daughter subjets of a jet, one expects that the reconstruction may be improved. Depending on the exact methodology used, these approaches have various names -- Jet Filtering\Cit{Butterworth:2008iy}, Pruning\Cit{Ellis:2009su}, Trimming\Cit{Krohn:2009th}, Cleansing\Cit{Krohn:2013lba}, Soft Drop\Cit{Larkoski:2014wba} and so on (Please see\Cit{Adams:2015hiv} and references therein for a more comprehensive listing and details). 
  
 A combination of Pruning in consort with a nondeterministic recombination jet algorithm was already explored in Ref.\Cit{Ellis:2012sn}, for the case of the boosted-$W^\pm$ topology. As mentioned earlier, our aim here is to explore, in some detail, the potential benefits obtained in combining the nondeterministic approach of Qjets with jet Trimming; taking  as prototypes the case of a diboson resonance topology. 
 
 In the case of Trimming, the contamination is reduced by applying a threshold cut on the transverse momentum  of each subjet, constituting the jet to be groomed. One requires each daughter subjet to satisfy,
\begin{equation}
p_{T} \ > \ f_\text{cut} \cdot \Lambda_\text{hard}\; .
\end{equation}
Here, $f_\text{cut}$ and $\Lambda_\text{hard}$ are parameters of the Trimming algorithm. For instance, one may choose $\Lambda_\text{hard}$ to correspond to the transverse momentum of the jet. Like most grooming methods, one of the crucial things to note is that significant gains are obtained because the backgrounds are attenuated drastically in some tagging window, relative to signal, as a consequence of grooming. Trimming, in going from algorithms designed for boosted heavy particles to one specifically designed for light parton jets, gives additional gains in topologies such as a dijet resonance\Cit{Krohn:2009th}. This latter feature is specifically interesting and is among the motivations for the present investigation.

It should also be commented that the Trimming parameters may be dynamically varied to suit the kinematics of an event and the exact values may be deduced iteratively from Monte-Carlo, or real data, tuned to standard candles. As emphasised in the introduction, numerous studies have now corroborated Trimming, across diverse scenarios at the LHC\Cit{Chatrchyan:2013vbb, LENZI:2013wta, GUETA:2013rta, LOCH:2014lla, CMS:kxa, Aad:2013gja, TheATLAScollaboration:2013ria}.

\section { \label{sec:3}Nondeterministic Approach to Jets and Jet Substructure}

We now review the idea of nondeterministically constructing tree histories, and how the procedure leads to greater significance, and provides new variable discriminants for signal characterization.

\subsection{\label{subsec:3.1}The Qjets Paradigm}

As we described in \Sec{sec:2}, jets arise out of the partons, through a series of soft and collinear radiation showers; clustered finally using some jet algorithm of choice. These showers are thought to evolve hierarchically, in decreasing order of transverse momenta or branching angles. An immediate issue that one must address then is whether the jet algorithm employed faithfully inverts the parton shower. The answer is generally in the negative. A given set of final state particles, in an event, may have evolved through a multitude of intermediate steps. Therefore, the intermediate $1\rightarrow 2$ branchings of the shower, leading to a tree structure, is not always well defined. This ambiguity may occasionally lead to different algorithms constructing a jet observable differently. Depending on the details of one's selection criteria, it may then occur that the construction from one of these algorithms fails the tagging requirement.
 
 This inherent non-invertible nature was the motivation behind considering a nondeterministic way to construct the trees\Cit{{Ellis:2012sn},{Ellis:2014eya}}. In this viewpoint, a \emph{set of trees} (with some appropriate weights) is associated with \emph{each jet}. In the earlier, deterministic way of constructing trees (which we shall refer to as the {\em conventional} method), all information about the ambiguity in clustering and eventual jet-tagging is discarded. Thus, for instance, in the conventional method, any two jets that pass a set of selection criteria are assigned the same weights, even if one is unambiguously tagged and the other only marginally satisfied the tagging criteria.  Thus, conventional tagging efficiency per jet is \textit{binary}; taking the value $1$ or $0$, depending on if the tagging criteria is satisfied by an individual jet or not.
 
 \begin{figure}
\centering
\includegraphics[scale=0.4]{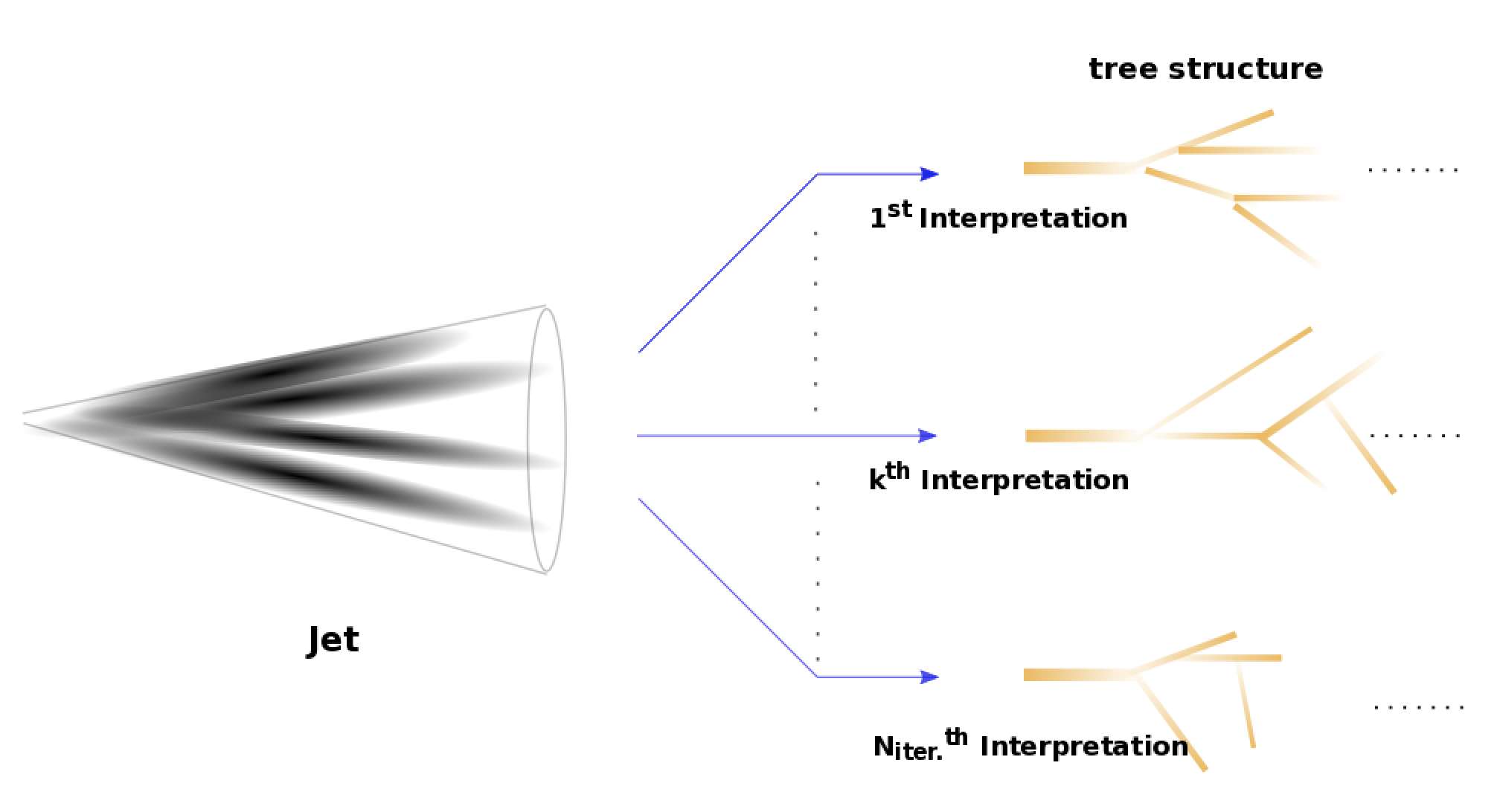}
\caption{Ambiguities in jet-algorithms may be mitigated by constructing a history of trees. A non deterministic approach to tree-based jet substructure, like Qjets, associates to each jet an appropriately weighted family of trees.}
\label{fig:dijet_invmass_sig_qjet}
\end{figure}

  The stochastic approach tries to rectify this by assigning a distribution of trees for each jet. This may equivalently be viewed, if desired, as processing a jet using a range of algorithms and grooming procedures. This concept is illustrated in \Fig{fig:dijet_invmass_sig_qjet}. In the  stochastic method, the  tagging efficiency exhibits a \textit{continuous} range of values, in the interval $\left[0,1 \right]$. This is a manifestation of the fact that with each jet we are now associating a distribution of appropriately weighted trees, and associated observables. The stochastic approach was shown to have two marked advantages\Cit{{Ellis:2012sn},{Ellis:2014eya}}
 
 \begin{itemize}
 
 \item An improvement in the statistical stability of jet observables and tagging. This was shown to provide an increase in discovery potential $S/\delta B$, and yield an effective luminosity gain as a consequence; the stochastic approach reduces the statistical fluctuation $\delta B$ for the same number of background jets relative to conventional clustering approaches.
 
\item One may now associate to each jet a \textit{distribution} for any observable of interest. This was termed \textit{volatility}, and was shown to have powerful discriminating characteristics, in distinguishing signal-like and QCD-like events (This was shown specifically in the case of the jet mass in the earlier studies, but the idea is more general)

\end{itemize}

Given a set of final-state four-momenta, associating a set of trees to each jet in an event  would be impractical if one had to consider all tree histories in its entirety. It is nevertheless seen that a good approximation to such a procedure, capturing all salient features of the tree distributions, can be achieved through a method akin to Monte-Carlo algorithms. Since IR and collinear safe observables are insensitive to small rearrangements in momenta, the set of trees deviating drastically from conventional clustering have very small weights, while the set of tress relatively closer to the conventional tree-history give very similar weights. This equips us to extract most of the required information by considering a small fraction of the trees, in a computationally efficient manner. As already mentioned, many aspects of this paradigm has already been validated at the LHC, by both the ATLAS and CMS collaborations\Cit{TheATLAScollaboration:2013tia,CMS:2013uea,Khachatryan:2014vla, CMS:2014joa}.

We will incorporate this method to find the subjets, of a given jet, for the purpose of Trimming. A conventional subjet finding algorithm, taken as a part of ordinary Trimming, assigns binary tagging probabilities to each subjet. This is meant in the sense that each subjet found is either discarded or kept, depending on its respective $p_T$. This could subsequently manifest itself, indirectly, as a binary tagging probability for the trimmed jet. With the nondeterministic algorithm, each jet gets assigned a history of subjet components. Each of these subjet interpretations is then taken as input into the Trimming procedure, and like in the earlier case, gives better statistical stability because the tagging probability for the parent jet now has a continuous distribution in the interval $\left[0,1 \right]$. In addition, as before, a mass-volatility variable can be applied effectively if the jet being trimmed has an intrinsic mass-scale; to characterise it from generic QCD background.

\subsection{\label{subsec:3.2} Qjet Observables and Statistics} 

In this subsection, we will define the  variables of interest and review in brief how the statistical robustness comes about. 

In the conventional Trimming case (henceforth denoted by a superscript `c' or `T'), for each jet (labelled by subscript $\alpha$) the tagging probability is binary, in that it is either tagged or not tagged
\be
\tau^c_\alpha = 
\left \{
\begin{aligned} 
1 ~ &~ \text{\small{Conventionally trimmed jet}} \in \text{\small{tagging criteria}} \;  \\
0 ~ &~ \text{\small{Conventionally trimmed jet}} \notin \text{\small{tagging criteria}}\\
 \end{aligned} \right. \; .
 \label{eq:3.1}
\ee
The tagging criteria could be some specific set of requirements; the simplest being a mass-window criterion, for instance. The total number of tagged jets (or equivalently events, if we are considering a single jet of interest in each event and its tagging efficiency is being directly taken as the tagging efficiency for the event) in this case will be
\be
N_\text{\tiny{tag}}^c =  \sum^{\text{N}}_{\alpha=1 }  \tau^c_\alpha \; ,
\ee
where $N$ is the total number of jets/events being considered. The corresponding fractional uncertainty in the number of tagged jets is given by 
\be
\left(\frac{\delta N_\text{\tiny{tag}}}{N_\text{\tiny{tag}}}\right)^c=\frac{1}{\sqrt{\epsilon^c_\text{\tiny{N}} N}}\;,
\ee
where $\epsilon^c_\text{\tiny{N}}$ denotes the conventional tagging efficiency
\be
\epsilon^c_\text{\tiny{N}} =\frac{N^c_\text{\tiny{tag}}}{N}\;.
\ee
 This is just the mean $\langle \tau^c \rangle$.

In the QTrimming case, each jet is being interpreted multiple times and therefore has a tagging probability in the interval $\left[0,1 \right]$. With the multiple interpretations, in combination with some grooming procedure, any jet characteristic will now also have a distribution of values. Let us consider details pertaining to the  tagging efficiencies first. We will use a superscript `q' or `QT' to denote the stochastic cases. 

Each jet (say, labelled by $\alpha$ again), now being interpreted $\text{N}_{\text{\tiny{iter}}}$ times (\Fig{fig:dijet_invmass_sig_qjet}) gives a tagging efficiency 
\be
\tau^q_\alpha =  \frac{1}{\text{N}_{\text{\tiny{iter}}}} \sum^{\text{N}_{\text{\tiny{iter}}}}_{k=1 }    
\left \{
\begin{aligned} 
1 ~ &~ k^\text{\tiny{th}}\,\text{\small{interpretation}} \in \text{\small{tagging criteria}} \;  \\
0 ~ &~k^\text{\tiny{th}}\,\text{\small{interpretation}} \notin \text{\small{tagging criteria}}\\
 \end{aligned} \right. \; .
 \label{eq:3.6}
\ee
The `number' of tagged jets is therefore
\be
N_\text{\tiny{tag}}^q =  \sum^{\text{N}}_{\alpha=1 }  \tau^q_\alpha \; .
\ee
In the nondeterministic case, it may be shown that one gets a non-trivial result for the corresponding fractional uncertainty in tagged jets\Cit{Ellis:2014eya}  
\be
\left(\frac{\delta N_\text{\tiny{tag}}}{N_\text{\tiny{tag}}}\right)^q=\frac{1}{\sqrt{N}}\times\sqrt{1+\frac{\sigma_{\tau^q}^2}{\langle \tau^q \rangle^2}}\; .
\ee

Here, $\langle \tau^q \rangle$ and $\sigma_{\tau^q}^2$ are the mean and variance respectively, of the $\tau^q_\alpha$ tagging efficiencies, computed over all the N events. In the nondeterministic approach, we may equate the mean signal efficiency ($\epsilon^q_\text{\tiny{S}}$) and mean background efficiency ($\epsilon^q_\text{\tiny{B}}$) to the corresponding $\langle \tau^q_\text{\tiny{S}} \rangle$ and  $\langle \tau^q_\text{\tiny{B}} \rangle$ respectively. Note that one may consider conventional tagging as the limit where the stochastic tagging efficiency $\tau^q_\alpha$, per jet, approaches either 0 or 1.

One of the crucial points to note is that for any probability distribution\Cit{Ellis:2014eya}, 
\be
\frac{1}{\sqrt{N}}~~\leq~\left(\frac{\delta N_\text{\tiny{tag}}}{N_\text{\tiny{tag}}}\right)\leq~\frac{1}{\sqrt{\langle \tau \rangle N}}\;,
\label{eq:3.8}
\ee
since $ \sigma_\tau^2 \leq \langle \tau \rangle (1-\langle \tau \rangle)$. This translates to a condition
\be
\frac{\delta N_\text{\tiny{tag}}}{\sqrt{N_\text{\tiny{tag}}}}~=~\sqrt{\frac{\langle \tau^2 \rangle}{\langle \tau \rangle}}~\leq~1\;,
\label{eq:3.9}
\ee

since $\langle \tau^2 \rangle \leq \langle \tau \rangle$. Note that the upper limit is saturated for the conventional case, where $\delta N_\text{\tiny{tag}} =\sqrt{N_\text{\tiny{tag}}}$. The upper-bound being saturated by the fractional uncertainty in conventional tagging implies that, compared to conventional tagging which is binary, Qjet tagging can lead to significant reductions in the fractional uncertainty in cross-section measurements.

Putting these ideas together, we can quantify the relative change in significance or discovery potential, between conventional and stochastic cases, as a figure-of-merit 
\be
\frac {\left(S/\delta B\right)^q}{\left(S/\delta B\right)^c}= \frac{\epsilon^q_\text{\tiny{S}}}{\epsilon^c_\text{\tiny{S}}}\cdot     \frac{1}{ \frac{\delta B^q}{\sqrt{B^q}}  \sqrt{\frac{\epsilon^q_\text{\tiny{B}}}{\epsilon^c_\text{\tiny{B}}}}}\;.
\label{eq:3.10}
\ee
Here, $S$ is the number of tagged signal events
\be
S^\text{c/q} =\sum_{\alpha\in\text{\tiny{signal}}} \tau_\alpha^\text{c/q}\; .
\label{eq:3.11}
\ee
Conventional tagging efficiencies, as mentioned earlier, may be taken as the limit where each $\tau_\alpha$ is either 0 or 1; compared to the Qjet case where $\tau_\alpha \in [0,1]$. $\delta B^\text{c/q}$ is the rms fluctuation in background events  that pass the tagging criteria,
 \be
 B^\text{c/q} =\sum_{\alpha\in\text{\tiny{bkg.}}} \tau_\alpha^\text{c/q}\;.
 \label{eq:3.12}
 \ee
 $\epsilon^\text{c/q}_\text{\tiny{S/B}}$ are the efficiencies as defined earlier, and identified with $\langle \tau_\text{\tiny{S/B}} \rangle$. Note that an equivalent methodology would be to use pseudoexperiments, to define the figure-of-merit, as implemented in the original study\Cit{Ellis:2012sn}.
 
The significance ($S/\delta B$) is proportional to the square-root of the luminosity ($\sqrt{\mathcal{L}}$), and any relative improvement in it (more specifically the square of $S/\delta B$) may be interpreted as an effective improvement in the luminosity. Another way to state this would be that if there is improvement through stochastic jet grooming, one may obtain the same discovery potential for a much reduced integrated luminosity.

Let us now turn to the fact that in the non-deterministic approach, with multiple interpretations of a single jet, one also obtains distributions for jet observables. This novel feature may now be utilised for signal-background discrimination, by making judicious cuts on variables constructed from these distributions. For instance, a \textit{mass-volatility}  \Cit{{Ellis:2012sn},{Ellis:2014eya}}, per jet, may be defined as
\be
\mathcal{V}_{m}=\Gamma_{m}/ m^\text{c} \; ,
\label{eq:mass-vol}
\ee
where $\Gamma_{m}=\sqrt{\text{Var.}(m)}=\sqrt{\langle m^2 \rangle-\langle m \rangle^2}$ is the rms deviation of the jet mass (over the $\text{N}_{\text{\tiny{iter.}}}$ interpretations), and $m^\text{c}$ is the conventional jet mass that would be obtained from a conventional jet algorithm. Note that in the earlier literature\Cit{{Ellis:2012sn},{Ellis:2014eya}}, the rms deviation was divided by the average mass $\langle m \rangle$, over $\text{N}_{\text{\tiny{iter.}}}$ interpretations, to get a dimensionless mass-volatility. Here we now use a slightly different normalisation, without loss of generality. In our investigation, the mass-volatility is calculated without imposing any signal-window criteria on the Qjets, but requires that the corresponding conventionally trimmed jet fall within a suitably defined signal mass-window. This allows for an honest comparison of the two methods. As has been pointed out \Cit{{Ellis:2012sn},{Ellis:2014eya}}, if the object on which the Qjet algorithm is being run has an intrinsic mass-scale, for instance boosted $W^{\pm}\rightarrow jj$ , one expects its mass-volatility to be generally lower than QCD jets (which usually have no intrinsic mass scales associated with them, in the limit we are working under.) This may be leveraged as we shall see to improve searches. As an aside, also note that this idea is generally applicable to any jet observable of interest that has a distribution, after applying the nondeterministic algorithm. For a general jet observable $\mathcal{O}$, a $\mathcal{O}$-volatility may be defined as 
\be
\mathcal{V}_{\mathcal{O}}=\Gamma_{\mathcal{O}}/\mathcal{O}^\text{c}\; .
\ee
Here, $\Gamma_{\mathcal{O}}=\sqrt{\text{Var.}(\mathcal{O})}$, over the $\text{N}_{\text{\tiny{iter.}}}$ interpretations, and $\mathcal{O}^\text{c}$ is the conventional value of the jet observable, that would again be obtained from a conventional jet algorithm. It is conceivable that a study of these $\mathcal{O}$-volatility distributions may further help in characterization; this nevertheless is beyond the scope of our present study.

In the next section we describe the procedure we follow, along with simulation details. In \Sec{sec:5} we will then demonstrate that one indeed gains significantly, in terms of an enhancement in discovery potential, through stochastic grooming.
\section{\label{sec:4} Stochastic Jet Grooming Methodology}

As pointed out in Sec.~\ref{sec:3}, the Qjets paradigm brings improvements through two novel features: $(i)$ Modified tagging efficiencies, leading to significantly improved statistical stability; and $(ii)$ New observables from distribution of jet variables, in the dimension of clustering histories. This in combination with Trimming -- QTrimming -- is the focus of our investigation. In this section we describe our procedure and the scenarios we consider, to illustrate its performance.

\subsection{\label{sec:4.1}The QTrimming Procedure}

Given an event, we identify and reconstruct objects using a suitable set of parameters; we will list the values adopted in \Sec{subsec:4.3}. With the reconstructed jets in hand, we apply QTrimming, as follows -- 

\begin{enumerate}
\item The input to QTrimming is a given jet, for example, an anti-$k_T$ jet with a given $p_T$ and $R$. The algorithm works with the \textit{constituents} of the jet. (Note that as far as QTrimming is concerned,  the algorithm does not really care whether the constituents are detector level output, such as tracks, calorimeter cells or particle flow, or simply hadrons from the shower.)
\item Calculate all pairwise distances $d_{ij}$, given by Cambridge-Aachen measure \Cit{Dokshitzer:1997in, Wobisch:1998wt}
\be
\label{eq:dij}
	d_{ij} \ = \  \Delta R_{ij}^2 \ = \  \left(y_i - y_j\right)^2 + \left(\phi_i - \phi_j \right)^2 \; , 
\ee
where, $\Delta R_{ij}$ is the angular distance between a pair of $4$-momenta. As usual,  $y$ and $\phi$ represent the rapidity and the azimuthal angle respectively.  

\item For each pair of $4$-vectors, calculate the probability of merging from the distance measures :
\be
\label{eq:omega}
P_{ij}^{(\alpha)} \ =\  \frac{1}{\mathcal{N}}  
	\exp\left\{  -\alpha \frac{(d_{ij}- d_\text{min})}{d_\text{min}} \right\}, 
   \qquad \text{where} \qquad 
    d_\text{min} \ = \ \text{Min} \left\{ d_{ij} \right\} \; .
\ee
Following Ref\Cit{Ellis:2012sn, Ellis:2014eya}, $\alpha$ and $\mathcal{N}$ denote the \textit{rigidity} parameter, and the normalization.

\item After a pair is chosen and merged, repeat steps 2-4, until all particles satisfy the following
\be
d_\text{min} \ \geq \ R_{\text{trim}}\; .
\ee
At this point, all the $4$-vectors remaining (call them protojets) are passed to the grooming algorithm. 

\item Discard all  protojets with $p_{T_i} < f_\text{cut} \cdot  p_{T}^J$. Here $f_\text{cut}$ is a fixed dimensionless parameter and $p_{T}^J$ is the transverse momentum of the original jet. The jet Trimming algorithm implemented here closely follows that of Ref.~\cite{Krohn:2009th}.  

\item Repeat steps 2-5, $N_\text{iter.}$ number of times, to arrive at $N_\text{iter.}$ different interpretations of the same jet. 
\end{enumerate}
 
 We apply this procedure to each signal and background jet that is to be QTrimmed. We will then impose a tagging criteria, that will be detailed in \Sec{sec:5}, to estimate the statistical significance and figure-of-merits. In the next subsection we describe our prototypical signal and background events. Details of the simulation, and the parametric values adopted, will be listed in \Sec{subsec:4.3}
 
\subsection{Signal Topologies Considered}
As discussed earlier, to make a detailed comparison we will consider the case of a diboson resonance -- considering separately the hypothetical cases where the daughter particles from the resonance decay are \textit{massless/light} or \textit{massive}. We do this, in the context of our comparison, by neglecting or utilising the mass-volatility discriminant in various ways. It will be shown that QTrimming as a tool smoothly migrates between both these cases, and yields enhanced discovery potential under all circumstances, albeit in slightly different ways. Proceeding in this fashion helps us to faithfully compare cases where the enhancement of signal significance is purely due to the added statistical robustness, versus cases where the significance is augmented due to volatility discriminants.

Consider the case of a diboson resonance ($\Phi$), produced in $pp$-collisions, decaying to $W^\pm$ bosons which subsequently decay hadronically 
\be
p~p \ \rightarrow \ \Phi \ \rightarrow \  W^\pm\left( q\bar{q}'\right)  W^\pm\left( q\bar{q}'\right) \;.
\ee

We take for the mass of the resonance $m_\Phi \ = \ 1\tev $. Since $m_W \ll m_\Phi$, the decay products of $W^\pm$ will be collimated due to the boost. The events are thus characterised by two back to back jets, say $J_L$ (Leading jet) and $J_{NL}$ (Next-leading jet), where each jet contains the all-hadronic decay products of a $W^\pm$-boson.

In a hypothetical scenario, we may speculate that the $\Phi$ will be discovered primarily through two searches: 
\begin{itemize}
\item A dijet resonance search -- here, no jet-mass information is used \textit{per se} and one hunts for a bump on top of the di-jet mass ($m_{J_L J_{NL}}$) continuum.  
\item A di-$W^\pm$ resonance search where one tries to find a bump on top of the di-jet mass continuum and each jet ($J_L$ and $J_{NL}$) is now also required to be tagged as a $W^\pm$.
\end{itemize}

The former is relevant to digluon/diquark resonances, where the two daughter particles from the resonance decay are light/massless and have no intrinsic mass scales, while the latter will also be of relevance to cases where such an intrinsic mass scale provides additional handles in the analysis. These are the two scenarios we shall consider. In both cases we will take QCD dijets as our main background.

\subsection{\label{subsec:4.3}Simulation Details and Parameters}
 
We use \texttt{Pythia 8.215}\Cit{Sjostrand:2014zea} to generate our signal (single production of $\Phi$ decaying to a pair of hadronically-decaying $W^\pm$ bosons), as well as background events (QCD-dijets). In all the cases, the signal events along with associated QCD-dijet background are generated at $13\,\rm{TeV}$-LHC. For the background events,  we impose a parton-level, transverse momentum cut $\hat{p}_T > 400\gev$. No additional cut is imposed for the signal events at the partonic level. PU has been modelled by over-laying minimum bias, soft-QCD events drawn from a poisson distribution, with mean $\langle N_{PU} \rangle$, over each signal/QCD-background events. We consider $\langle N_{PU} \rangle= 40$ for the comparisons, thereby hoping to give conservative values for the improvements. 

\texttt{Delphes 3.3.2} is used to simulate the details of the  ATLAS detector, after the addition of PU (For further details, see~\cite{deFavereau:2013fsa}). We only collect the tower outputs from Delphes, and additional functionalities such as jet reconstruction or energy rescaling of Delphes are not used in our study. At the tower level, all entries with $p_T < 1\gev$ associated with the hadronic calorimeter are discarded. 

The Delphes outputs are then clustered into jets by  \texttt{FastJet 3.1.3}\Cit{{hep-ph/0512210},{Cacciari:2011ma}}.  We use anti-$k_T$ algorithm\Cit{Cacciari:2008gp} with parameters $R=0.7$ and $p_{T}^\text{\tiny{min}} = 500\gev$. Events are selected as long as it contains at least two jets. It is required that the two leading jets (in $p_T$) are found in the central part of the detector (\textit{i.e.} $\left| \eta \right| \leq 2.0 $).  From each selected event, the two leading jets (let us call them $J_L$ and $J_{NL}$ respectively) are retained for further analyses. In all the analyses, a simple mass window requirement is taken as the tagging criteria. The relevant mass windows for the resonance $\Phi$ and $W^\pm$, where applicable, will be taken as $\Omega_\Phi\equiv \left(  1000 \pm \sqrt{1000}\right)\gev$ and $\Omega_W \equiv \left(  65-95 \right)\gev$. The jet Trimming parameters are chosen as the optimised values, from the conventional analyses, as will be detailed in Sec.~\ref{sec:5}. We take $R_{\text{trim}}=0.2$ and $f_\text{cut}=0.03$, which are used in all the analyses. For the QTrimming analyses we will take $N_\text{iter}=100$, which may be demonstrated to be sufficient to capture all relevant Qjet characteristics. We will also consider various values of the rigidity parameter, $\alpha$, for comparisons and to validate limits.

Let us now consider the various cases of interest, the results from the analyses and how they interpolate among each other, displaying the gains from QTrimming in the process.

\section{\label{sec:5} Results} 
In the context of discovering a hypothetical $\Phi$ resonance, we shall consider four different approaches. In each case, a given event will be analysed in two different ways -- a conventional analysis and a QTrimming analysis. We will progressively add more information from QTrimming in each subsequent analysis to clarify the gains obtained by considering various aspects. This will also provide a nice picture of how the cases interpolate smoothly among each other, depending on the analyses.

\subsection{Dijet Resonance Search (Analysis I) }
Consider the first case. Let the conventional and QTrimming analyses for this case be defined as follows :

\begin{itemize}
 \item[\circlew{dj}]  \textit{Conventional Analysis}:  Trim the leading jet ($J_L$) and next-leading jet ($J_{NL}$) in the usual way\Cit{Krohn:2009th} using the set of optimized trimming parameters. Let us call each of the Trimmed jets $J_L^c$ and $J_{NL}^c$. The combination of the two conventionally trimmed jets is a provisional candidate for $\Phi$,
 
 \be
 J_L^c \oplus J_{NL}^c \equiv J_\Phi^c \; .
 \ee
 
 We will consider $J^c_\Phi$ to be tagged as coming from $\Phi$ decay if the jet-mass falls within the mass window $\Omega_\Phi$ (defined in \ref{subsec:4.3})
 \be
 m_{J_\Phi^c} \in \Omega_\Phi~\Rightarrow~\text{Tagged}\;.
 \ee
 This obviously gives a binary probability (i.e. $0$ or $1$), say $\tau^c_{J_\Phi-\Phi}$. Equate $\tau^c_{J_\Phi-\Phi}$ to \textit{tagging an event }as $\Phi$-like
 
 \be
\tau_{\text{\tiny event}}^{\text {\tiny c, I}}\equiv \tau^c_{J_\Phi-\Phi}\; .
\label{eq:5.3}
\ee

To reiterate, the right hand side is the conventional tagging probability of the reconstructed dijet, defined in accordance with \Eq{eq:3.1}, and this is being taken as the conventional tagging probability for the event as a whole.

\begin{figure}[h]
\centering
\includegraphics[scale=0.575]{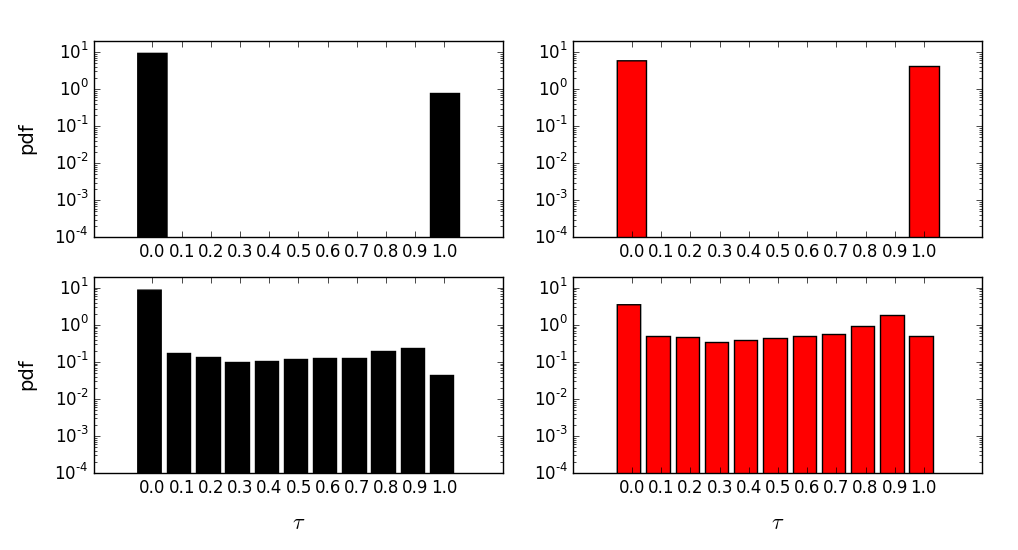}
\caption{\label{fig:ex1} Event tagging efficiencies in Analysis I ($\tau_{\text{\tiny event}}^{\text {\tiny I}}$), for QCD background (Left, Black) and signal (Right, Red). The tagging efficiencies for the conventional case  (Top figures) , $ \tau^c_{J_\Phi-\Phi}$, and stochastic case (Bottom figures), $ \tau^q_{J_\Phi-\Phi}$, are as defined in \Eq{eq:5.3} and \Eq{eq:5.6}. The stochastic QTrimming case is for $\alpha=1.0$. The figure is area normalized, with a bin size of $0.1$. }
\end{figure}

  \item[\circleb{dj}] \textit{QTrimming Analysis}: Trim $J_L$ and $J_{NL}$ now using the procedure detailed in \ref{sec:4.1}, with $N_\text{iter.}=100$; using the same optimized trimming parameters as in the conventional case above. In each interpretation (say $k$-th, as in \Fig{fig:dijet_invmass_sig_qjet}) we obtain two trimmed jets -- $J_L^k$ and $J_{NL}^k$. Each of the interpretations now furnish a provisional candidate for $\Phi$
 
 \be
 J_\Phi^k  \equiv  J_L^k \oplus J_{NL}^k ~~~~~;~~~k \in [1,N_\text{iter.}]\; .
 \ee 
  
We now therefore have a distribution of $ N_\text{iter}$ masses -- $\{m_{J_\Phi^k } \} $, as discussed in \ref{subsec:3.1}. In this case, the probability that the dijet will be tagged as a $\Phi$, following \Eq{eq:3.6}, is
\begin{equation}
\tau^q_{J_\Phi-\Phi} \ \equiv \ \frac{1}{N_\text{iter}} \sum_{k=1}^{N_\text{iter}} \left\{ 
	\begin{aligned}   
		1 \quad  &  m_{J_\Phi^k } \in \Omega_\Phi \\
		0 \quad  &  m_{J_\Phi^k } \not\in \Omega_\Phi  
	\end{aligned} \right. \; ,
\label{eq:5.5}	
\end{equation}
and this is equated with the probability that the \textit{event will be tagged} as $\Phi$-like
\be
\tau_{\text{\tiny event}}^{\text {\tiny q, I}}\equiv \tau^q_{J_\Phi-\Phi}\; .
\label{eq:5.6}
\ee
Unlike the binary event probability, $\tau_{\text{\tiny event}}^{\text {\tiny c, I}}$, in the conventional analysis, this is some number in the interval $[0,1]$ as we mentioned in \Sec{subsec:3.2}. We may now use this event tagging efficiency from QTrimming, $\tau_{\text{\tiny event}}^{\text {\tiny q, I}}$, to quantify how $\Phi$-like the reconstructed event is. We can then directly use the formulas, from \Sec{subsec:3.2}, to calculate signal and background efficiencies and the variables of interest.  

 \end{itemize}
 
 The event tagging efficiencies for background and signal events are shown in \Fig{fig:ex1}, in the conventional and stochastic cases. Here, following \Eq{eq:5.3} and \Eq{eq:5.6}, the conventional and stochastic event tagging efficiencies (Top and botton figures respectively) are illustrated, for the background QCD (Left figures, Black) and signal dijet events (Right figures, Red). The figure is shown with $\alpha=1.0$, in the QTrimming case. The figure is also area normalised, with a bin size of $0.1$, as expected of a probability density function (pdf).
 
\begin{table}[h]
\begin{center}
\begin{tabular}{| c | c  c | c | c  c | }
  \hline
  $\alpha$ & $ S^{\text{QT}} / S^\text{T}   $ 
 & $ B^{\text{QT}} / B^\text{T}   $  
& $ \frac{\left(\delta B/\sqrt{B} \right)^\text{QT}}{\left(\delta B/\sqrt{B} \right)^\text{T}}    $ 
& $\frac{\left(S/ B \right)^\text{QT}}{\left( S/ B \right)^\text{T}} $ 
& $\frac{\left(S/\delta B \right)^\text{QT}}{\left( S/\delta B \right)^\text{T}} $\\
  \hline
  \hline
  $10.0$  & $1.12$ & $1.07$ & $0.93$ & $1.04 $ & $1.16$ \\
  $1.0$  & $1.10$ & $1.07$ & $0.85$ & $1.02 $ & $1.24$ \\
  $0.1$  & $0.78$ & $0.88$ & $0.79$ & $0.89 $ & $1.05$ \\
  $0.01$  & $0.37$ & $0.56$ & $0.78$ & $0.66 $ & $0.63$ \\ 
  \hline
\end{tabular}
\end{center}
\caption{\label{table:1} Results for Analysis I, relevant to diquark/digluon resonance searches, for various values of the rigidity parameter $\alpha$. Here, as mentioned before, we have taken $\langle N_{\text{PU}} \rangle = 40$ for the comparison. T and QT superscripts refer to the conventional and QTrimming cases in Analysis I respectively. The gains from statistical robustness is exemplified by the fact that one obtains an $S/\delta B$ ratio greater than unity for various rigidity parameter choices.}
\end{table}
 
 With these definitions we may calculate the various quantities of interest. The results for the full analysis are presented in Table.\ref{table:1}, for various choices of the rigidity parameter $\alpha$, as defined in Eq.~(\ref{eq:omega}). For brevity, we have only shown in the table those quantities explicitly relevant for the computation; of signal-background separations and discovery potentials. As $\alpha\rightarrow \infty$, the QTrimming procedure should approach the conventional Trimming procedure, and for higher values of the rigidity, much greater than those shown in the table, this is indeed observed. This expected behaviour has already been pointed out in \Cit{Ellis:2012sn}. 
 
  As $\alpha$ is reduced from very high values (i.e. the conventional Trimming limit), the binomial-like tagging efficiency distributions become more spread out. Some of the events, in which conventional trimming yielded a dijet mass out of the mass window $\Omega_\Phi$ (i.e. with $\tau_{\text{\tiny event}}^{\text {\tiny c, I}}=0$), now have after QTrimming $\tau_{\text{\tiny event}}^{\text {\tiny q, I}} > 0$. This is simply due to the fact that in some of the interpretations, the QTrimmed masses for the dijets now lie inside the mass window. Similarly, a fraction of the events conventionally classified by $\tau_{\text{\tiny event}}^{\text {\tiny c, I}}=1$, now also have interpretations outside the mass window $\Omega_\Phi$, after QTrimming, and hence move to bins with $\tau_{\text{\tiny event}}^{\text {\tiny q, I}} < 1$. 
  
For rigidity choices $\alpha \gtrsim 0.1$, the number of interpretations moving into the mass window is seen to be greater than those moving out, for both signal and background events. This is reflected in the fact that  $ S^{\text{QT}} / S^\text{T} > 1$ and $ B^{\text{QT}} / B^\text{T} > 1$ for these values in Table.\ref{table:1}. The difference in the specific trends has to do with the fact that the signal has a dijet mass distribution peaking around $\Omega_\Phi$ while the QCD background is a falling distribution. This is subtly visible, for instance, in \Fig{fig:ex1} with the signal events (Lower-right, red) having a slightly higher weight near $\tau=1.0$, compared to background events (Lower-left, black).
 
 Note also that, as motivated in \Eq{eq:3.8} and \Eq{eq:3.9}, one indeed finds $\delta N_\text{\tiny{tag}}/\sqrt{N_\text{\tiny{tag}}} < 1$ for the stochastic case, yielding better statistical stability for both signal and background. This is shown explicitly in Table.\ref{table:1}, for the $\delta B/\sqrt{B}$ ratios; note that $\delta B_\text{\tiny{tag}}=\sqrt{B_\text{\tiny{tag}}}$ in the conventional Trimming case (T), and hence the denominator is just unity. The trend in the ratios of signal-background separation ($S/B$) is understood directly from the behaviour of $S^{\text{QT}} / S^\text{T}$ and $B^{\text{QT}} / B^\text{T}$, and the fact that the signal has a dijet mass distribution peaking near $\Omega_\Phi$ while the QCD background has a falling distribution as mentioned earlier.
 
  Under the current analysis, the gains are purely from the statistics of Qjets. The improvement in discovery potential $S/\delta B$ is predominantly from two sources. Following \Eq{eq:3.10}, note that
  \be
  \frac{\left(S/\delta B \right)^\text{QT}}{\left( S/\delta B \right)^\text{T}} = S^{\text{QT}} / S^\text{T}   \cdot  \frac{\delta B^\text{T}}{\delta B^\text{QT}}= S^{\text{QT}} / S^\text{T}   \cdot  \sqrt{\frac{B^\text{T}}{B^\text{QT}}}\cdot \left( \frac{\delta B^\text{T}/\sqrt{B^\text{T}}}{\delta B^\text{QT}/\sqrt{B^\text{QT}}} \right)\; .
  \ee
  
 So, the $S/\delta B$ ratios are a function of three quantities. Owing to the square root, the quantity $\sqrt{\frac{B^\text{T}}{B^\text{QT}}}$ is close to unity in most of the relevant cases (See Table.\ref{table:1}). Therefore, most of the interesting variations are from the other two competing terms. A part of the contribution is from improvement in $S^{\text{QT}} / S^\text{T}$, and a part is from the improvement in the $\left( \frac{\delta B^\text{QT}/\sqrt{B^\text{QT}}}{\delta B^\text{T}/\sqrt{B^\text{T}}} \right)$ ratio. They are both decreasing, with a decrease in the rigidity parameter, and there is an interplay now between the two statistical contributions. One observes the best improvement of about $\mathcal{O}(25\%)$, for $\alpha=1.0$. A relative improvement in  $S/\delta B$ is equivalent to an effective luminosity gain, as we mentioned earlier. 

 Note that the scenario we have considered in analysis I is relevant for any general dijet resonance searches, where the individual daughter jets ($J_L$ and $J_{NL}$) have no implicit mass scales per se. In the next subsection we will consider a general case where this is not so and investigate how similar procedures affect signal-background separation and signal significance.
 
\subsection{Di-$W^\pm$ Resonance Search (Analysis II, III, \& IV)}

Consider the case of QTrimming in the context of discovering $\Phi$ in a di-$W$ resonance search. As we previously motivated, the concepts discussed in the di-$W^\pm$ resonance context are also relevant to any search where the daughter jets have an intrinsic mass-scale -- di-$Z^0$, di-Higgs, di-top resonances etc. The conventional analysis remains almost the same, as in Analysis I, with the addition of conventional W-tagging, but based on how we construct the QTrimming event tagging efficiencies, we will distinguish three subclasses -- which we shall call Analysis II, III and IV. Let us begin by defining the common aspects of the analyses :

\begin{itemize}

 \item[\circlew{dw}] \textit{Conventional Analysis}: As in the earlier analysis we trim $J_L$ and $J_{NL}$, using the optimized trimming parameters (call them $J_L^c$ and $J_{NL}^c$ as before). We consider 
$J_L^c$ and $J_{NL}^c$ as W-tagged if  the respective jet masses fall within the W-mass window -- $m_{J_L^c}, m_{J_{NL}^c} \in \Omega_W$. $ \Omega_W$ is as defined in \ref{subsec:4.3}.
The combination of the the two W-tagged jets, $J_\Phi^c  \equiv  J_L^c \oplus J_{NL}^c $, is a provisional candidate for $\Phi$. We define $J^c_\Phi$ to be tagged, as a $\Phi$-resonance, if it falls within the $\Phi$ mass window $\Omega_\Phi$ (i.e. $m_{J_\Phi^c} \in \Omega_\Phi$). For such a tagged event, the corresponding conventional tagging efficiency, $\tau^c_{J_\Phi-\Phi}$, is of course 1. Let the full event tagging efficiency be now defined as 

 \be
\tau_{\text{\tiny event}}^{\text {\tiny c, II/III/IV}}\equiv \tau^c_{J_\Phi-\Phi}\times \tau^c_{J_{L}-W } \times \tau^c_{J_{NL}-W } 
\label{eq:5.7}
\ee

\begin{figure}[h]
\centering
\includegraphics[scale=0.575]{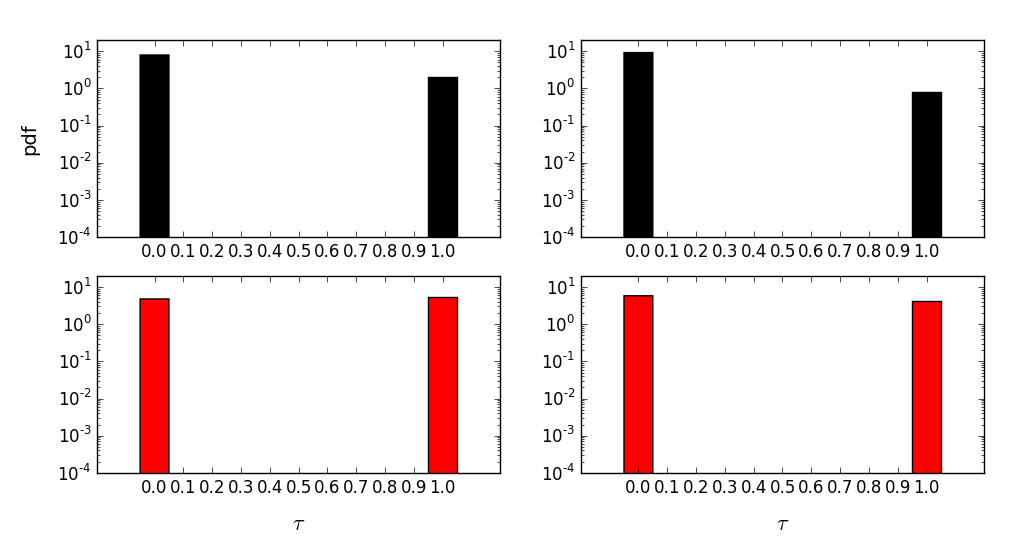}
\caption{\label{fig:Ex2_C_WTag_LNL} The bimodal, conventional W-tagging efficiency distributions for background (Top, Black) and signal (Bottom, Red) for the leading jet $J_{L}$ (Left) and next-leading jet $J_{NL}$ (Right).}
\end{figure}

The conventional W-tagging efficiencies ($ \tau^c_{J_{L}-W }$ and $\tau^c_{J_{NL}-W }$) are defined as 

\begin{equation}
\tau^c_{J_L-W } \ =   \left\{ 
	\begin{aligned}   
		1 \quad  &  m_{J_L^c } \in \Omega_W \\
		0 \quad  &  \text{Otherwise}  
	\end{aligned} \right. \qquad 
\tau^c_{J_{NL}-W } \ =   \left\{ 
	\begin{aligned}   
		1 \quad  &  m_{J_{NL}^c } \in \Omega_W \\
		0 \quad  &  \text{Otherwise}  
	\end{aligned} \right. \; .
\end{equation}

Their distributions are shown in \Fig{fig:Ex2_C_WTag_LNL}, for QCD dijet background (Top figures, Black) and diboson signal (Bottom figures, Red) events. The left and right figures correspond to the leading ($J_L$) and next-leading ($J_{NL}$) jets respectively. In the conventional case, the tagging efficiencies are binary for each event and this is clear from the bimodal distribution observed. The conventional $\Phi$-tagging efficiency distribution, $\tau_{\text{\tiny event}}^{\text {\tiny c, II/III/IV}}$, is illustrated by the top figures of \Fig{fig:Ex2_CQ_ETag}, for QCD (Top-Left) and signal events (Top-right) respectively.

 \item[\circleb{dw}] \textit{QTrimming Analysis}:   Trim $J_L$ and $J_{NL}$ using the same optimized trimming parameters, but using the algorithm of \ref{sec:4.1}, again with $N_\text{iter} = 100$. In a given run (say, $k$-th) we again obtain two trimmed jets ($J_L^k$ and $J_{NL}^k$ ) and a candidate for $\Phi$, (the dijet $J_\Phi^k  \equiv  J_L^k \oplus J_{NL}^i $). As explained before, we obtain three sets of $ N_\text{iter}$ masses  $\{  m_{J_L^k }  \} $, $\{  m_{J_{NL}^k }  \} $, and $\{  m_{J_\Phi^k }  \} $.

\end{itemize} 
 
Let us now differentiate three ways we may proceed depending on how the stochastic event tagging efficiencies may be constructed. The idea will be to progressively include more information from QTrimming as we proceed from one analysis to the other.

\subsubsection{Analysis II}
One choice is to define the stochastic event tagging probability as  
\begin{equation}
\tau_{\text{\tiny event}}^{\text {\tiny q, II}} \equiv \ \tau^q_{J_\Phi-\Phi} \times  \tau^c_{J_{L}-W } \times \tau^c_{J_{NL}-W } 
\label{eq:5.9}
\end{equation}

\begin{figure}[h]
\centering
\includegraphics[scale=0.575]{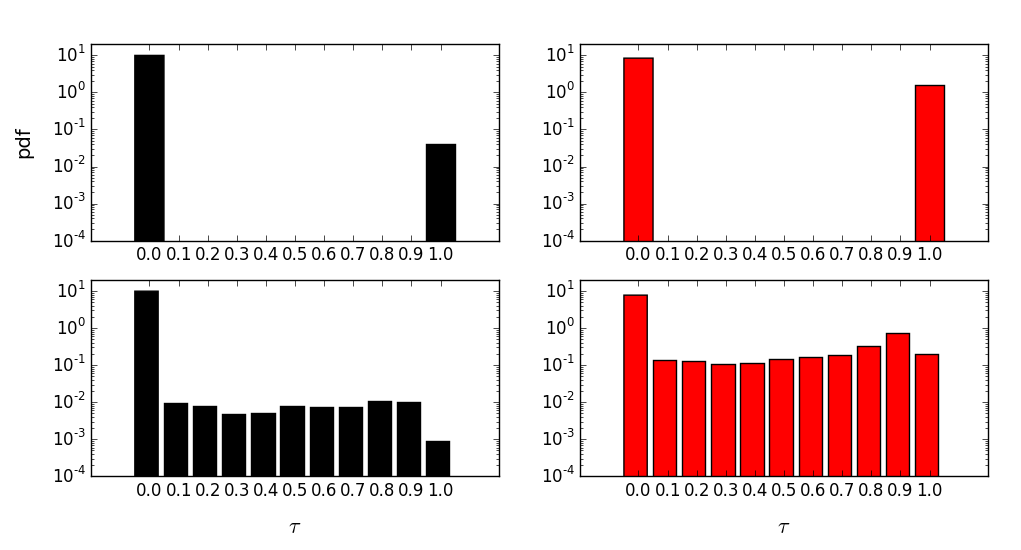}
\caption{\label{fig:Ex2_CQ_ETag} The QCD background (Left figures, Black) and diboson signal (Right figures, Red) event efficiencies for Analysis II ($\tau_{\text{\tiny event}}^{\text {\tiny II}}$). The top figures are for the conventional case, as defined by \Eq{eq:5.7}, and the bottom ones for the stochastic case, as defined by \Eq{eq:5.9}. The stochastic case is for $\alpha=1.0$ and being pdfs the figures are area normalised as before. }
\end{figure}

Here, $\tau^q_{J_\Phi-\Phi}$ is defined as in \Eq{eq:5.5} and $\tau^c_{J_{L}-W } \times \tau^c_{J_{NL}-W }$  are the products of the conventional W-tagging efficiencies (See \Fig{fig:Ex2_C_WTag_LNL}). Note that this choice incorporates information only from  conventional W-tagging ($\tau^c_{J_{L}-W }$ and $\tau^c_{J_{NL}-W }$) along with the QTrimming $\Phi$-tagging efficiency ($\tau^q_{J_\Phi-\Phi}$) -- in this sense it is the simplest way to incorporate some minimal information from QTrimming, while using a W-tag requirement on the daughter jets. The respective event tagging efficiency distributions are displayed in \Fig{fig:Ex2_CQ_ETag} (Bottom figures). Iterating over all the events, the various quantities may again be computed, and the final results are displayed in \Tab{table:2}. The general trends are again very similar to those discussed in \Tab{table:1}. As $\alpha$ is reduced from values corresponding to the conventional limit ($\alpha\rightarrow \infty$), the binomial-like tagging efficiency distribution again spreads out corresponding to various interpretations. The trends in signal and background are again slightly different owing to the nature of their dijet and jet mass distributions near $\Omega_\Phi$ and $\Omega_W$. 

With the addition of conventional W-tagging, one observes improvements in $\delta N_\text{\tiny{tag}}/\sqrt{N_\text{\tiny{tag}}}$ and $S/\delta B$ for the QTrimming case here. The improvement in $S/\delta B$ with respect to the conventional case is again from an interplay between improvements in $S^{\text{QT}} / S^\text{T}$ and $\delta B/\sqrt{B}$ ratios. The highest gain in $S/\delta B$, with the addition of conventional W-tagging to QTrimming $\Phi$-tagging efficiency, is of the order of $20\%$, for $\alpha=1.0$.

\begin{table}[h]
\begin{center}
\begin{tabular}{| c | c  c | c | c c |}
  \hline
  $\alpha$ & $ S^{\text{QT}} / S^\text{T}   $ 
  & $ B^{\text{QT}} / B^\text{T}   $  
& $ \frac{\left(\delta B/\sqrt{B} \right)^\text{QT}}{\left(\delta B/\sqrt{B} \right)^\text{T}}    $ 
& $\frac{\left(S/ B \right)^\text{QT}}{\left( S/ B \right)^\text{T}} $ 
& $\frac{\left(S/\delta B \right)^\text{QT}}{\left( S/\delta B \right)^\text{T}} $\\
  \hline
  \hline
  $10.0$  & $1.06$ & $1.07$ & $0.92$ & $1.04 $ & $1.16$ \\
  $1.0$  & $1.03$ & $1.07$ & $0.84$ & $0.96 $ & $1.19$ \\
  $0.1$  & $0.77$ & $0.87$ & $0.78$ & $0.89 $ & $1.06$ \\
  $0.01$  & $0.39$ & $0.54$ & $0.77$ & $0.72 $ & $0.69$ \\
  \hline
\end{tabular}
\end{center}
\caption{\label{table:2} Results from Analysis II, for $\langle N_{\text{PU}} \rangle = 40$. The analysis uses a combination of conventional W-tagging information with $\Phi$-tagging information, from QTrimming, as quantified in \Eq{eq:5.5}  }
\end{table}

\subsubsection{Analysis III}
Let us proceed by incorporating more information from Qjets. In keeping with this perspective, let use now use information from QTrimming in $W$-tagging as well. Define the probabilities for $W$-tagging $J_L$ and $J_{NL}$, after QTrimming, as 
\begin{equation}
\tau^q_{J_{L}-W } \ \equiv \ \frac{1}{N_\text{iter}} \sum_{k=1}^{N_\text{iter}} \left\{ 
	\begin{aligned}   
		1 \quad  &  m_{J_L^k } \in \Omega_W \\
		0 \quad  &  m_{J_L^k } \not\in \Omega_W  
	\end{aligned} \right. \qquad 
\tau^q_{J_{NL}-W } \ \equiv \ \frac{1}{N_\text{iter}} \sum_{i=1}^{N_\text{iter}} \left\{ 
	\begin{aligned}   
		1 \quad  &  m_{J_{NL}^k } \in \Omega_W \\
		0 \quad  &  m_{J_{NL}^k} \not\in \Omega_W  
	\end{aligned} \right. \; .
\end{equation}

\begin{figure}[h]
\centering
\includegraphics[scale=0.575]{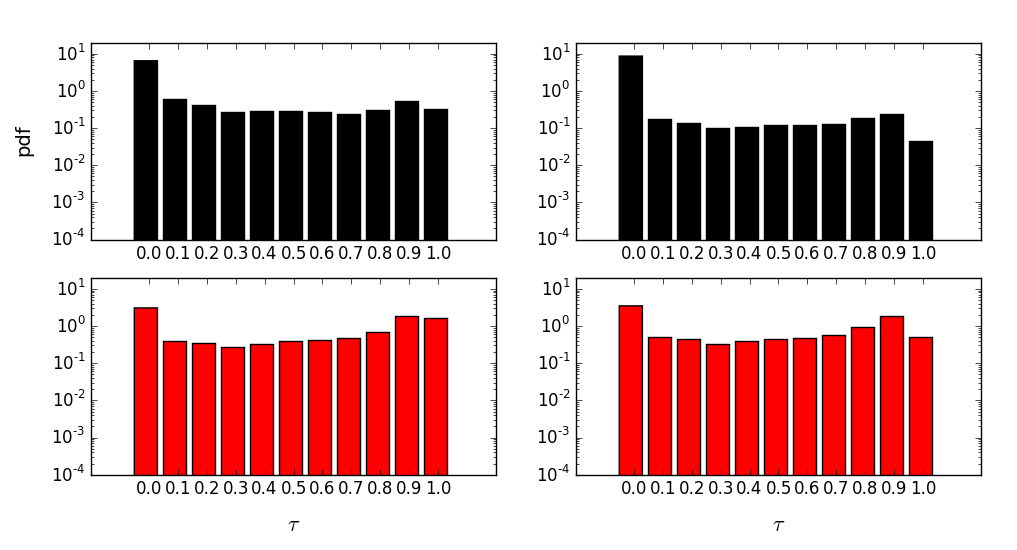}
\caption{\label{fig:Ex3_Q_WTag_LNL} Stochastic W-tagging efficiencies, with $\alpha=1.0$, for the leading (Left figures) and next-leading jets (Right figures) in QCD background (Top figures; black) and diboson signal events (Bottom figures; red).}
\end{figure}

The stochastic W-tagging distributions are as shown in \Fig{fig:Ex3_Q_WTag_LNL}, for the leading and next-leading jet. The figure is shown as before for $\alpha=1.0$ and $\langle N_{\text{PU}} \rangle = 40$, with area normalisation. The distribution is markedly different from the bimodal distribution of \Fig{fig:Ex2_C_WTag_LNL}, for conventional W-tagging. For different choices of $\alpha$ there is now a non-trivial behaviour in various interpretations leaking into or leaking out of the $\Omega_W$ mass window.

\begin{figure}[h]
\centering
\includegraphics[scale=0.575]{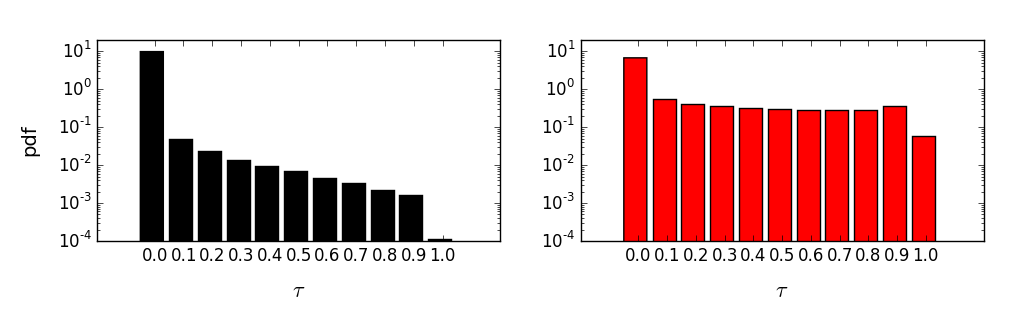}
\caption{\label{fig:Ex3_Q_ETag} The $\alpha=1.0$, stochastic event tagging efficiency distribution $\tau_{\text{\tiny event}}^{\text {\tiny q, III}}$ for QCD background (Left; black) and diboson signal (Right; red). These are also incorporating information from stochastic W-tagging as expositioned in  \Eq{eq:5.11}.}
\end{figure}

Using these definitions, a total event tagging probability pertinent to the present case may be defined as
\begin{equation}
\tau_{\text{\tiny event}}^{\text {\tiny q, III}} \ \equiv \ \tau^q_{J_\Phi-\Phi} \times  \tau^q_{J_{L}-W } \times \tau^q_{J_{NL}-W } \; .
\label{eq:5.11}
\end{equation}
This event tagging efficiency exhibits a distribution as shown in \Fig{fig:Ex3_Q_ETag}. Note the drastic change in event tagging efficiencies between \Fig{fig:Ex2_CQ_ETag} and \Fig{fig:Ex3_Q_ETag}. Incorporating the stochastic W-tagging efficiency with the QTrimming $\Phi$-tagging efficiency has improved overall signal-event tagging relative to background, compared to the previous analysis, and yields better statistical stability for $\delta B/\sqrt{B}$.

The computed results are shown in Table~\ref{table:3}. The results represent the maximum gain one can obtain utilising only Qjet statistics (i.e. without using any information from volatility variables). The statistical robustness as quantified by $\delta N_\text{\tiny{tag}}/\sqrt{N_\text{\tiny{tag}}}$ is also improved compared to Analysis II (Table~\ref{table:2}). The relative gain in signal significance is found to be as much as $70\%$ for a value of $\alpha=1.0$. 

\begin{table}[h]
\begin{center}
\begin{tabular}{| c | c  c | c | c  c|}
  \hline
  $\alpha$ & $ S^{\text{QT}} / S^\text{T}   $ 
  	& $ B^{\text{QT}} / B^\text{T}   $  
		& $ \frac{\left(\delta B/\sqrt{B} \right)^\text{QT}}{\left(\delta B/\sqrt{B} \right)^\text{T}}    $ 
			& $\frac{\left(S/ B \right)^\text{QT}}{\left( S/ B \right)^\text{T}} $ 
				& $\frac{\left(S/\delta B \right)^\text{QT}}{\left( S/\delta B \right)^\text{T}} $\\
  \hline
  \hline
 $10.0$  & $1.16$ & $0.97$ & $0.82$ & $1.20 $ & $1.44$ \\
 $1.0$  & $1.07$ & $1.07$ & $0.61$ & $1.00 $ & $1.71$ \\
 $0.1$  & $0.48$ & $1.90$ & $0.46$ & $0.25 $ & $0.77$ \\
 $0.01$  & $0.11$ & $2.03$ & $0.51$ & $0.05 $ & $0.15$ \\
  \hline
\end{tabular}
\end{center}
\caption{\label{table:3} Results for Analysis III. The analysis utilizes tagging probabilities for $J_\Phi$, $J_L$ and $J_{NL}$ from QTrimming. The results represent the best possible gains that may be obtained solely from the statistical robustness provided by QTrimming. Enhanced statistical robustness, compared to Analysis I (Table~\ref{table:1}) and Analysis II (Table~\ref{table:2}), as well as significant gains in $S/\delta B$ are observed. }
\end{table}

\subsubsection{Analysis IV}

Finally, let us include both tagging and mass-volatility information furnished by QTrimming. This case will represent the use of a maximal set of information, provided by QTrimming, in the context of our specific methodology. Unlike a diquark or digluon resonance case, where the daughter jets have no implicit mass-scales and mass-volatility is not useful, here it is useful for tagging $W^\pm$ bosons.

Motivated by the presence of an implicit mass-scale, let us define the event tagging efficiency in this case to be
\begin{equation}
\tau_{\text{\tiny event}}^{\text {\tiny q, IV}} \ \equiv \ \tau^q_{J_\Phi-\Phi} \times  \tau^q_{J_{L}-W } \times \tau^q_{J_{NL}-W }  \times  \left\{ 
	\begin{aligned}   
		1 \quad  &  \mathcal{V}^{J_L }_m~\text{and}~\mathcal{V}^{J_{NL} }_m < ~\mathcal{V}^\text{cut}_m \\
		0 \quad  &  \text{Otherwise}  
	\end{aligned} \right. \; ,
\label{eq:5.12}	
\end{equation}
thereby including all the tagging information from QTrimming as well as the mass-volatility information of $J_L$ and $J_{NL}$. The mass-volatility $ \mathcal{V}_m$ is as defined in \Eq{eq:mass-vol}. In this case the pure tagging efficiencies are similar to the distributions in \Fig{fig:Ex3_Q_ETag}, with an added cut on the mass-volatilities of the leading and next-leading jets. 

\begin{figure}[h]
\centering
\includegraphics[scale=0.575]{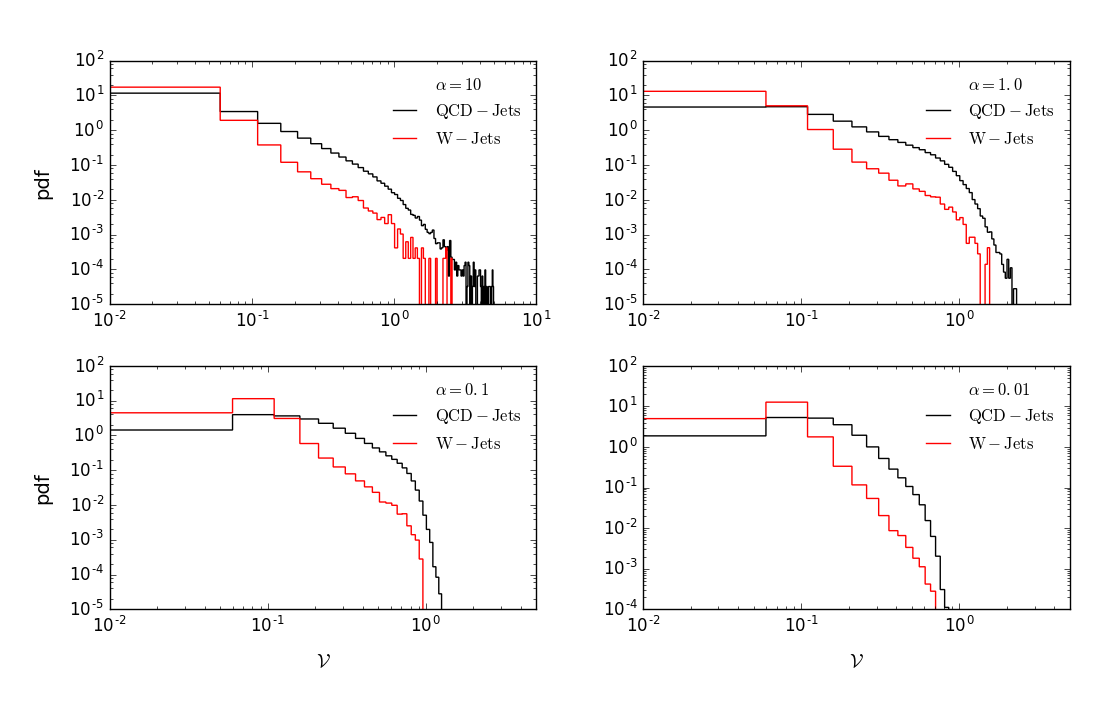}
\caption{\label{fig:vol} Mass-volatility distributions for hadronically decaying, boosted $W^\pm$ jets and QCD-jets, for various values of the rigidity parameter $\alpha$. As in all the analyses, we have assumed $\langle N_\text{PU} \rangle = 40$. Observe that generically the W-jets have a smaller mass-volatility after QTrimming compared to QCD jets. Placing an additional cut with respect to the mass-volatilities may be expected to therefore add to the event tagging efficiencies. This expectation is validated by Table~\ref{table:4}.}
\end{figure}

The volatility cut preferentially selects W-like jets, which have an intrinsic mass-scale. The mass-volatility distributions for the signal and QCD jets are as shown in \Fig{fig:vol} for comparison, with  $\langle N_\text{PU} \rangle = 40$ and for various values of the rigidity parameter $\alpha$. Note from the figure that on average the W-jets have lower mass-volatilities than QCD jets. The final results for the $S/B$ and $S/\delta B$ ratios are shown in Table~\ref{table:4} for various choices of $\mathcal{V}_\text{cut}$ and $\alpha$. The cuts may be optimised in a realistic analysis. The various quantities have been computed, as before, using the conventional and stochastic event tagging efficiencies defined by \Eq{eq:5.7} and \Eq{eq:5.12}. For brevity, we have only shown the final results for the signal-background separation and discovery potential ratios.

It is seen that with the inclusion of all the tagging and mass-volatility information, from QTrimming, one obtains drastic improvements in both signal-background separation (namely, $S/B$) and discovery potential (namely, $S/\delta B$). With the $\mathcal{V}_\text{cut}$, in addition to the stochastic tagging efficiencies, the number of signal events successfully tagged relative to background events increases significantly. There is now an interplay here between the nature of the dijet/jet mass distributions near the relevant mass-windows as well as the presence or absence of an intrinsic mass-scale, depending on whether it is a $W^\pm$ jet or a QCD jet. This is reflected in Table~\ref{table:4}, by the large ratios in $S/B$ compared to conventional event tagging. The discovery potential $S/\delta B$ also improves greatly compared to all the previous analyses. This is again sourced by improvements in $S^{\text{QT}} / S^\text{T}$ and $\delta B/\sqrt{B}$, though now from both Qjet statistics and mass-volatility cuts.

\begin{table}[h]
\begin{center}
\begin{tabular}{| c| |c|c|c|c | c|c|c|c| }
 \hline
  \multirow{5}{*}{$\mathcal{V}_\text{cut} $} 
  	&  \multicolumn{4} { c |}{ \multirow{2}{*}{ $\left(S/ B \right)^\text{QT}_{\mathcal{V} < \mathcal{V}_\text{cut}  }
		 / \left( S/ B \right)^\text{T} $ }  } 
		 &  \multicolumn{4} { | c | }{ \multirow{2}{*}
		 	{ $\left(S/ \delta B \right)^\text{QT}_{\mathcal{V} < \mathcal{V}_\text{cut}  }
		 		/ \left( S/ \delta B \right)^\text{T} $ }  }    \\  
	& \multicolumn{4} { c | }{ \multirow{4}{*}{  } } & \multicolumn{4} { |c| }{ \multirow{4}{*}{  } }  \\ 
		\cline{2-5} \cline{6-9}
   &  \multicolumn{4} { c| }{$\alpha $ }  &  \multicolumn{4} { c |}{$\alpha $ }  \\  
   &  $10.0$ & $1.0$ & $0.1$  & $0.01$ &  $10.0$ & $1.0$ & $0.1$  & $0.01$  \\
  \hline
   $0.50$ &  $1.20$ & $1.00$ & $0.26$  & $0.05$ &  $1.44$ & $1.71$ & $0.77$  & $0.15$  \\
   $0.25$ &  $1.22$ & $1.12$ & $0.36$  & $0.06$ &  $1.45$ & $1.74$ & $0.83$  & $0.15$  \\
   $0.10$ &  $1.49$ & $2.67$ & $5.82$  & $1.87$ &  $1.51$ & $2.12$ & $1.62$  & $0.57$  \\
   $0.09$ &  $1.57$ & $3.28$ & $7.82$  & $3.17$ &  $1.52$ & $2.20$ & $1.49$  & $0.59$  \\
   $0.08$ &  $1.65$ & $4.22$ & $9.96$  & $8.41$ &  $1.53$ & $2.30$ & $1.16$  & $0.80$  \\
   $0.07$ &  $1.78$ & $5.51$ & $16.9$  & $8.96$ &  $1.55$ & $2.35$ & $1.00$  & $0.55$  \\
  \hline   

\end{tabular}
\end{center}
\caption{\label{table:4}  Results for Analysis IV incorporating the full tagging and mass-volatility information, from QTrimming. Substantial gains are obtained in this case, as seen from the table, with signal-background separation and discovery potentials enhanced drastically in many cases. This validates the notion that QTrimming has the potential to improve and augment searches.}
\end{table}

\pagebreak

\section {\label{sec:6} Summary and Conclusions}

Jet Trimming has already been shown to greatly improve event reconstruction in hadron collisions by mitigating contamination due initial state radiation, multiple interactions, and event pileup. On the other hand the Qjet paradigm is a powerful technique to decrease random statistical variations of observables, yielding large boosts in discovery potential. It also provides new hitherto unavailable distributions that may be used to further discriminate signal-like and background-like events. 

In this work we considered a combination of the jet grooming method of Trimming in tandem with Qjets; the subjets to be trimmed were constructed from the constituents of the jet in a stochastic manner. We investigated the case of a heavy resonance decaying into boosted, hadronically decaying $W^\pm$. Various cases were considered, by progressively incorporating more information from the stochastic approach. This enabled us to consider various subtleties in how gains were obtained and in smoothly interpolating between cases where the improvements were purely from statistical robustness and those where it was reinforced by volatility variable discriminants. 

In all cases it was shown that the discovery potentials can be greatly increased through a combination of Trimming with Qjet methods. Although demonstrated explicitly for a particular grooming method and event topology, some of the general ideas in our exposition is expected to have wide reaching applicability. We hope to explore some of these possibilities in a forthcoming work.

  %

\section*{Acknowledgements}
T.S.R and A.T would like to acknowledge discussions with D. Krohn and L.T. Wang during an earlier investigation that partly motivated aspects of the present work. A significant part of the computations were performed on the Mapache cluster in the HPC facility at LANL. T.S.R would like to thank MITP (Mainz, Germany) and A.T wishes to thank the DTP group at TIFR (Mumbai, India) for hospitality, during the completion of this work. This work was supported in part by DOE grants DOE-SC0010008 and a SERB Early Career Award.

\bibliography{QTrimming}
\bibliographystyle{jhep}

\end{document}